\documentclass[aps,prd,12pt,notitlepage,showpacs,nofootinbib,tightenlines]{revtex4-1}
\usepackage{amsmath}
\usepackage{amssymb}
\usepackage{epsfig}
\usepackage{graphicx}
\usepackage{bm}
\usepackage{times}
\usepackage{braket}
\usepackage{color}
\usepackage{slashed}
\usepackage{hyperref}

\definecolor{Red}{rgb}{1.,0.,0.}

\definecolor{Blue}{rgb}{0.,0.,1.}

\definecolor{nicered}{rgb}{0.7,0.1,0.1}
\definecolor{nicegreen}{rgb}{0.1,0.5,0.1}
\hypersetup{colorlinks,citecolor=nicegreen,linkcolor=nicered}

\begin{document}
%%%%%%%%%%%%%%%%%%%%%%%%%%%%%%%%%%%%%%%%%%%%%

\newcommand{\beq}{\begin{eqnarray}}
\newcommand{\eeq}{\end{eqnarray}}
\newcommand{\ben}{\begin{enumerate}}
\newcommand{\een}{\end{enumerate}}
\newcommand{\be}{\begin{equation}}
\newcommand{\ee}{\end{equation}}

\newcommand{\non}{\nonumber\\ }

\newcommand{\jpsi}{J/\Psi}

\newcommand{\ppa}{\phi_\pi^{\rm A}}
\newcommand{\ppp}{\phi_\pi^{\rm P}}
\newcommand{\ppt}{\phi_\pi^{\rm T}}
\newcommand{\ov}{ \overline }

\newcommand{\zerot}{ {\textbf 0_{\rm T}} }
\newcommand{\kt}{k_{\rm T} }
\newcommand{\fb}{f_{\rm B} }
\newcommand{\fk}{f_{\rm K} }
\newcommand{\rk}{r_{\rm K} }
\newcommand{\mb}{m_{\rm B} }
\newcommand{\mw}{m_{\rm W} }
\newcommand{\im}{{\rm Im} }

\newcommand{\pvsl}{ p \hspace{-2.0truemm}/_{K^*} }
\newcommand{\esl}{ \epsilon \hspace{-2.1truemm}/ }
\newcommand{\psl}{ p \hspace{-2truemm}/ }
\newcommand{\ksl}{ k \hspace{-2.2truemm}/ }
\newcommand{\lsl}{ l \hspace{-2.2truemm}/ }
\newcommand{\nsl}{ n \hspace{-2.2truemm}/ }
\newcommand{\vsl}{ v \hspace{-2.2truemm}/ }
\newcommand{\epsl}{\epsilon \hspace{-1.8truemm}/\,  }
\newcommand{\bfkk}{{\bf k} }
\newcommand{\calm}{ {\cal M} }
\newcommand{\calh}{ {\cal H} }
\newcommand{\calo}{ {\cal O} }

%%---------------------------------------------------------

%%
\title{Next-to-leading-order time-like rho electromagnetic form factors in ${k_{\rm T}}$ factorization }
\author{Ya-Lan Zhang$^{1}$}
\author{Si-Yu Wang$^{1}$}
\author{Chao Wang$^{1}$}
\author{Zhen-Jun Xiao$^{2}$}
\author{Ya Li$^{3}$}\email{Corresponding author:liyakelly@163.com}
\affiliation{1. Faculty of Mathematics and Physics, Huaiyin Institute of Technology,Huaian,Jiangsu 223200, P.R. China}
\affiliation{2. Department of Physics and Institute of Theoretical Physics, Nanjing Normal University, Nanjing, Jiangsu 210023, P.R. China}
\affiliation{3. Department of Physics, College of Sciences, Nanjing Agricultural University, Nanjing, Jiangsu 210095, P.R. China}
\date{\today}
%%
%%
%\vspace{1cm}
\begin{abstract}
We calculate the time-like $\rho$ electromagnetic (EM) form factor in the $k_T$ factorization formalism by including the next-to-leading-order (NLO) corrections of the leading-twist and sub-leading twist contributions.
It's observed that the NLO correction to the magnitude of the LO leading-twist form factor is lower than $30\%$ at large invariant mass squared $Q^2 > 30 \text{GeV}^2$.
It is found that the $\rho$ meson EM form factor is dominated by twist-3 contribution instead of by twist-2 one because of the end-point enhancement.
The theoretical predictions of the moduli of three helicity amplitudes and total cross section are analyzed at $\sqrt{s}=10.58$ GeV, which are consistent with measurements from BABAR Collaboration.

\end{abstract}

%\vspace{1cm}

\maketitle

\section{Introduction}
The EM form factor of the hadron plays a key role in understanding the hadron structures and the transition from the perturbative to the non-perturbative region.
It is an important physical observable which and has triggered both theoretical and experimental interests for a long time.

On the theoretical side, the $\rho$-meson EM form factors have been studied by models
based on the light-front formalism~\cite{Brodsky:1992px,Keister:1993mg,Cardarelli:1994yq,deMelo:1997hh}, the QCD sum rules~\cite{Braguta:2004kx}, and the light cone sum rules~\cite{Aliev:2004uj}, the light-front quark model~\cite{Choi:2004ww}, lattice calculation~\cite{Andersen:1996qb} and so on.
The cross-sections and various
polarization observables were calculated in terms of the EM form factors on basis of the model-independent formalism~\cite{Gakh:2006rj,Adamuscin:2007dt}.
A modified perturbative QCD (PQCD) approach based on the $k_T$ factorization \cite{Brodsky:1973kr,Collins:1991ty,Botts:1989kf,Li:1992nu,Huang:1989rd} has been applied to next-to-leading-order (NLO) analysis of several space- and time-like form factors, such as the pion EM form factor \cite{Hu:2012cp,Li:2010nn,Cheng:2015qra,Chen:2009sd,Cheng:2014gba},  the $\pi$-$\rho$ transition form factor~\cite{Hua:2018kho}, and the $B \rightarrow \pi$ transition form factors~\cite{Li:2012nk}.
For the $\rho$-meson~\cite{Zhang:2018bhj}, the NLO radiative
corrections to the space-like EM form factors with leading and sub-leading twist have also been studied in PQCD approach at at small invariant mass squared.
Recently, the BABAR Collaboration \cite{BaBar:2008fsh} extracted the helicity electromagnetic amplitudes for the $\gamma^*  \to \rho^+ + \rho^-$ from $e^+ + e^-$ annihilation producing the pair of $\rho$ mesons at $\sqrt{s}=10.58$~GeV.
The data brought information on the time-like $\rho$-meson form factors in the high-energy region where the relevance with the PQCD may be investigated.

The time-like region, accessible through annihilation reactions, is expected to investigate form factors which carries the information of momentum redistribution between all the constituents in initial and final states.
The most simple reaction which contains information on time-like $\rho$-meson form factors is the annihilation of an electron-positron pair into a $\rho^+\rho^-$ pair.
A particle of spin $S$ is characterized by $2S+1$ EM form factors.
For a spin-1 $\rho$-meson, there are three independent form factors as functions of momentum transfer squared $Q^2$, i.e. charge monopole ($G_1$), magnetic dipole ($G_2$) and  charge quadrupole ($G_3$), reads
    \begin{eqnarray} \label{3ffratio}
    	G_1:G_2:G_3=(1-\frac{2}{3}\eta):2:-1,
    \end{eqnarray}
    where is known as ``universal ratios" \cite{Brodsky:1992px}.
    The term $\eta=Q^2/4m^2_{\rho}$.
    The well-known asymptotic relation $G_1 \approx \frac{2\eta}{3}G_3$ \cite{Carlson:1984wr} is consistent with Eq. (\ref{3ffratio}) at high momentum transfer $\eta >>1$.

    In the space-like region, the universal ratio from PQCD was checked in the zero-binding energy limit of spin-1 composite particle \cite{Brodsky:1992px}.
    Sub-leading corrections to helicity matrix elements of the EM current beyond the universal leading PQCD have been considered in \cite{deMelo:2016lwr}, which results in the deviation of the "universal ratio" with the BABAR experiment.
   It may imply that either helicity conservation does not apply or
other reaction mechanisms contribute to the $\rho$ production.
   Therefore, the detailed analysis on sub-leading contributions to the $\rho$-meson EM form factors play the key role on checking the so-called ``universal ratio".
   Besides, the introduction of two-meson wave functions \cite{Diehl:1998dk}, whose parametrization also includes time-like form factors associated with different currents, has been required by the PQCD formalism for three-body $B$ meson decays.
A theoretical framework for three-body $B$ meson decays can be built if PQCD results for complex time-like form factors are reliable.
It is necessary to study the time-like EM form factor in PQCD approach.

   In this work, we consider the NLO corrections to the time-like form factors with the leading and sub-leading twists in the PQCD approach.
   With appropriate analytic continuation, NLO corrections to time-like form factors can be readily obtained from those to space-like ones from $-Q^2$ to $Q^2$.
   We calculate the three independent amplitudes to check the ``universal ratios" and estimate the cross section.
As the measurement of deuteron form factors in time-like region is beyond the present experimental possibilities, it is interesting to measure the EM form factors.

This paper is organized as follows. In the following section, we summarize the relevant kinetics and the input meson wave functions.
In Sec.~III, the NLO factorization formula for the rho meson time-like electromagnetic form factors will be calculated from the space-like one by analytical continuation.
Numerical results are performed in Sec.~IV.
And Sec.~V contains the conclusion.

\section{KINETICS AND INPUT WAVE FUNCTIONS}
\begin{figure}[htbp]
  \centering
  \vspace{0cm}
  \begin{center}
  \epsfxsize=10cm\epsffile{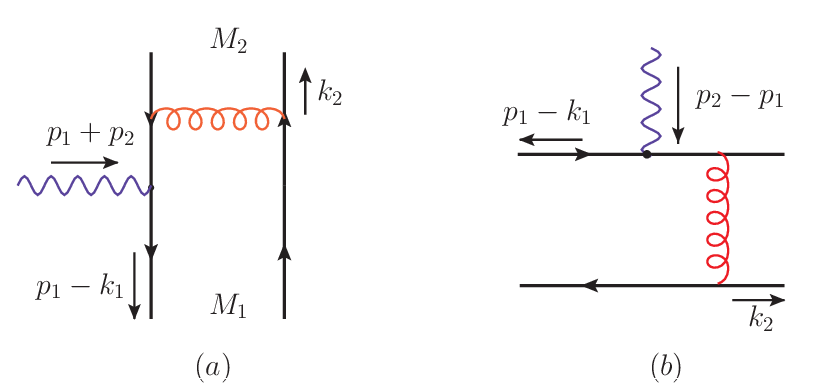}
  \end{center}
  \vspace{-0.8cm}
\caption{Feynman diagrams for time-like (a) and space-like (b) $\rho$ EM form factor at leading order, with $\bullet$ representing the EM vertex.}
\label{fig:time}
\end{figure}

In this section, we introduce the relevant kinetics and the input meson wave functions.
The LO quark diagrams for relative time-like and space-like $\rho$ EM form factors corresponding to the process $\gamma^{\ast}\rightarrow \rho\rho$ ($\rho\gamma^{\ast}\rightarrow \rho$) are displayed in Figs.~\ref{fig:time}(a) and ~\ref{fig:time}(b).
In the light-cone coordinate, the kinematic variables for meson  $\rho^-$ and $\rho^+$ in Fig.~\ref{fig:time}(a) are defined as
\begin{equation}
    P_1=\frac{Q}{\sqrt{2}}(1,\gamma_{\rho}^2,\textbf{0}_\textbf{T}),\quad    P_2=\frac{Q}{\sqrt{2}}(\gamma_{\rho}^2,1,\textbf{0}_\textbf{T}),
\label{eq:p1}
\end{equation}
where the dimensionless parameter $\gamma_{\rho}$ is defined as $\gamma_{\rho}^2\equiv m_{\rho}^2/Q^2$, with $m_{\rho}=0.77$ GeV being the mass of the $\rho$ meson, $Q^2=(p_1+p_2)^2$ being the invariant mass squared of the intermediate virtual photon.
The momentum $k_i(i=1,2)$ of the antiquark for $\rho^+(\rho^-)$ meson are of the form of
\begin{align}
k_1=(x_1\frac{Q}{\sqrt{2}},0, \textbf{k}_\textbf{{1T}}),\quad   k_2=(0,x_2\frac{Q}{\sqrt{2}},\textbf{k}_\textbf{{2T}}),
\label{eq:k1}
\end{align}
where the $x_i$ is antiquark momentum fraction and runs from zero to unity.
$\textbf{k}_\textbf{iT}$ is assigned to the transversal momentum.
The longitudinal and transversal polarization four-vectors of the $\rho^+(\rho^-)$ meson in the final state are defined as
\begin{align}
\epsilon_{1\mu}(L)=\frac{1}{\sqrt{2}\gamma_{\rho}}(1,-\gamma_{\rho}^2,\textbf{0}_\textbf{T}),\quad
\epsilon_{1\mu}(T)=(0,0,\textbf{1}_\textbf{T}),\nonumber\\
\epsilon_{2\mu}(L)=\frac{1}{\sqrt{2}\gamma_{\rho}}(-\gamma_{\rho}^2,1,\textbf{0}_\textbf{T}),\quad
\epsilon_{2\mu}(T)=(0,0,\textbf{1}_\textbf{T}).
\label{eq:2}
\end{align}
The relations $P_1\cdot \epsilon_1=0$, $P_2\cdot \epsilon_2=0$ and $\epsilon_1^2=\epsilon_2^2=-1$ in Eq.~(\ref{eq:p1}) and Eq.~(\ref{eq:2}) are satisfied.
The related expansions of the nonlocal matrix elements for $\rho$ meson are written in the following form \cite{Ball:1998sk,Ball:1998ff}
\begin{eqnarray}
&&\left\langle0\mid\bar{u}(0)_j d(z_1)_l\mid\rho^-(p_1,\epsilon_T)\right\rangle\nonumber\\
&=&\frac{1}{\sqrt{2N_c}}\int_0^1dx_1e^{ix_1p_1 \cdot z_1} \left\{\psl_1\epsl_{1\emph{T}}\phi_{\rho}^T(x_1)
+m_{\rho}\epsl_{1\emph{T}}\phi_{\rho}^v(x_1)
+m_{\rho}i\epsilon_{\mu\nu\rho\sigma}\gamma^{\mu}\gamma_5\epsilon_{1\emph{T}}^{\nu}n^{\rho}v^{\sigma}\phi_{\rho}^a(x_1)\right\}_{lj},\nonumber\\
&&\left\langle0\mid\bar{u}(0)_j d(z_1)_l\mid\rho^-(p_1,\epsilon_L)\right\rangle\nonumber\\
&=&\frac{1}{\sqrt{2N_c}}\int_0^1dx_1e^{ix_1p_1 \cdot z_1}\left\{m_{\rho}\epsl_{1\emph{L}}\phi_{\rho}(x_1)
+\psl_1\epsl_{1\emph{L}}\phi_{\rho}^t(x_1)
+m_{\rho}\phi_{\rho}^s(x_1)\right\}_{lj},\nonumber\\
&&\left\langle0\mid u(z_2)_j \bar{d}(0)_l\mid\rho^+(p_2,\epsilon_T)\right\rangle\nonumber\\
&=&\frac{1}{\sqrt{2N_c}}\int_0^1dx_2e^{ix_2p_2 \cdot z_2}\left\{\psl_2\epsl_{2\emph{T}}\phi_{\rho}^T(x_2)
+m_{\rho}\epsl_{2\emph{T}}\phi_{\rho}^v(x_2)
+m_{\rho}i\epsilon_{\mu\nu\rho\sigma}\gamma^{\mu}\gamma_5\epsilon_{2\emph{T}}^{\nu}v^{\rho}n^{\sigma}\phi_{\rho}^a(x_2)\right\}_{lj},\nonumber\\
&&\left\langle0\mid u(z_2)_j \bar{d}(0)_l\mid\rho^+(p_2,\epsilon_L)\right\rangle\nonumber\\
&=&\frac{1}{\sqrt{2N_c}}\int_0^1dx_2e^{ix_2p_2 \cdot z_2}\left\{ m_{\rho}\epsl_{2\emph{L}}\phi_{\rho}(x_2)
+\psl_2\epsl_{2\emph{L}}\phi_{\rho}^t(x_2)
+m_{\rho}\phi_{\rho}^s(x_2) \right\}_{lj},
\label{eq:6}
\end{eqnarray}
where $\phi_{\rho}$ and $\phi_{\rho}^T$ ($\phi_{\rho}^{v,a}$ and $\phi_{\rho}^{s,t}$) represent the twist-2 (twist-3) components for the distribution amplitudes (DAs) of $\rho$ meson, with $n=(1,0,\textbf{0}_\textbf{T})$ and $v=(0,1,\textbf{0}_\textbf{T})$ being dimensionless unit vectors in the light cone coordinate, and $N_C=3$ being the number of colors.

The explicit expressions of various twist DAs can be described as below \cite{Ball:1998sk,Chai:2022ptk}
\begin{eqnarray}
\phi_{\rho}(x) &=& \frac{3f_{\rho}}{\sqrt{6}} x(1-x) \left[1+a^{\|}_{2\rho}\frac{3}{2}(5t^2-1)\right],\\
\label{eq:rho-t2-1}
\phi^{T}_{\rho}(x)&=& \frac{3f^{T}_{\rho}}{\sqrt{6}} x(1-x) \left[1+a^{\bot}_{2\rho}\frac{3}{2}(5t^2-1)\right],\\
\label{eq:rho-t2-2}
\phi^{s}_{\rho}(x)&=& \frac{3f^{T}_{\rho}}{\sqrt{6}}(1-2x)\left[1 + a_{2\rho}^{\perp}(5t^2-1)\right],\\
\label{eq:rho-t3-s}
\phi^{t}_{\rho}(x)&=& \frac{f^{T}_{\rho}}{2\sqrt{6}}\left[3t^{2}+a_{2\rho}^{\perp}\frac{3}{2}t^2(5t^2-3)\right],\\
\label{eq:rho-t3-t}
\phi^{a}_{\rho}(x)&=&\frac{3f_{\rho}}{\sqrt{6}} (1-2x)\left[1+a_{2\rho}^{\parallel}(5t^2-1)\right],\\
\label{eq:rho-t3-a}
\phi^{v}_{\rho}(x)&=&\frac{f_{\rho}}{2\sqrt{6}}\left[\frac{3}{4}(1+t^{2})+a_{2\rho}^{\parallel}\frac{3}{7}(3t^2-1)
+a_{2\rho}^{\parallel}\frac{9}{112}(3-30t^2+35t^4)\right],
\label{eq:rho-t3-v}
\end{eqnarray}
with $t=2x-1$ and the Gegenbauer moments $a_{2\rho}^\parallel=0.17$ and $a_{2\rho}^{\perp} = 0.14$ \cite{Bharucha:2015bzk}.
The values of the longitudinal and transverse decay constants are taken as $f_\rho = 0.216$GeV, $f_\rho^T=0.165$GeV \cite{Ball:2006eu}.

\section{Next-to-leading order correction to time-like rho electromagnetic form factor}
In this section, we present the NLO factorization formulas for the $\rho$ meson time-like EM form factors in the framework of $k_T$ dependent factorization, where the contributions from the leading and sub-leading twist DAs of $\rho$ meson are considered.
Three Lorentz invariant time-like form factors $G_i(Q^2)(i=1,2,3)$, which are functions of time-like momentum transfers ($Q^2>0$), are formulated as the matrix element:
\begin{eqnarray}
\left\langle\rho^+(p_2,\epsilon_2^*)\rho^-(p_1,\epsilon_1)\mid J_{\mu,\mid \lambda \mid}\mid0\right\rangle
&=&-\epsilon_{2\beta}\epsilon_{1\alpha} \left\{ \left[(p_1+p_2)_{\mu}g^{\alpha\beta}-p_1^{\beta}g^{\alpha\mu}-p_2^{\alpha}g^{\beta\mu}\right]\cdot G_1(Q^2) \right. \nonumber\\
&&-\left[p_1^{\beta}g^{\mu\alpha}+p_2^{\alpha}g^{\mu\beta}\right]\cdot G_2(Q^2) \nonumber\\
&&\left.+\frac{1}{m_{\rho}^2}p_1^{\beta}p_2^{\alpha}(p_1+p_2)_{\mu}\cdot G_3(Q^2) \right\},
\label{eq:7}
\end{eqnarray}
where the $J_{\mu,\mid \lambda \mid}$ is the EM current of the spin-one particle.
According to the angular momentum conservation, there are three independent helicity amplitudes $F_{\rm LL}$, $F_{\rm LT/TL}$, and $F_{\rm TT}$ where the
indices indicate the helicities of the mesons $\rho$.
These independent helicity amplitudes are in terms of the linear combinations of three covariant form factors, which are represented as matrix elements of the hadronic EM current in the space-like regions and also in the time-like regions.
The relations are listed as follows \cite{Zhang:2015mxa}
\begin{eqnarray} \label{fform}
F_{\rm TT}(Q^2)&=&G_{1}(Q^2),\nonumber\\
F_{\rm LT/TL}(Q^2)&=&\frac{Q}{2m_{\rho}}\left[G_{1}(Q^2)+G_{2}(Q^2)\right],\nonumber\\
F_{\rm LL}(Q^2)&=&G_{1}(Q^2)-\frac{Q^2}{2m_{\rho}^2}G_{2}(Q^2)+\frac{Q^2}{m_{\rho}^2}\left(1+\frac{Q^2}{4m_{\rho}^2}\right)G_{3}(Q^2).
\end{eqnarray}

By introducing the transversal momentum $k_T$ to cancel the end-point singularity, integrating over the longitudinal momentum fractions and transversal coordinate conjugated to transversal momentum, one finds the PQCD predictions for the $Q^2$ dependence of the $\rho$ meson LO space-like EM form factors:
\begin{eqnarray}
F_{\rm LL}(Q^2)&=&-\frac{32}{3}\pi C_F Q^2 \alpha_s(\mu)\int_0^1dx_1dx_2 \int_0^{\infty}b_1db_1b_2db_2 \cdot \exp[-S_{\rho}(x_i;b_i;Q;\mu)]\nonumber\\
      && \Bigg\{ \left[-x_1+\frac{1}{2}\gamma_{\rho}^2(1-x_1)\right]\phi_{\rho}(x_1)\phi_{\rho}(x_2) \nonumber\\
       &&-\gamma_{\rho}^2\phi_{\rho}^t(x_1)\phi_{\rho}^s(x_2)-2\gamma_{\rho}^2(1+x_1)\phi_{\rho}^s(x_1)\phi_{\rho}^s(x_2) \Bigg\}\cdot h(x_1,x_2,b_1,b_2).
\label{hloll}
\end{eqnarray}
Similarly, the other two helicity amplitudes can be expressed as,
\begin{eqnarray}
F_{\rm LT}(Q^2)&=&-\frac{32}{3}\pi C_F Q m_{\rho} \alpha_s(\mu)\int_0^1dx_1dx_2 \int_0^{\infty}b_1db_1b_2db_2 \cdot \exp[-S_{\rho}(x_i;b_i;Q;\mu)]\nonumber\\
      && \Bigg\{ \frac{1}{2}\phi_{\rho}(x_1)\left[\phi_{\rho}^v(x_2)+\phi_{\rho}^a(x_2)\right]
      +\phi_{\rho}^T(x_1)\phi_{\rho}^s(x_2)\nonumber\\
       &&+\frac{1}{2}x_1[\phi_{\rho}^v(x_1)+\phi_{\rho}^a(x_1)]\phi_{\rho}(x_2) \Bigg\}\cdot h(x_1,x_2,b_1,b_2),
\label{hlolt}
\end{eqnarray}
\begin{eqnarray}
F_{\rm TT}(Q^2)&=&\frac{32}{3}\pi C_F  m_{\rho}^2 \alpha_s(\mu)\int_0^1dx_1dx_2 \int_0^{\infty}b_1db_1b_2db_2 \cdot \exp[-S_{\rho}(x_i;b_i;Q;\mu)] \nonumber\\
      &&\left\{(1+x_1)\left[\phi_{\rho}^a(x_1)\phi_{\rho}^a(x_2)-\phi_{\rho}^v(x_1)\phi_{\rho}^v(x_2)\right]\right.\nonumber\\
       &&\left.+(1-x_1)\left[\phi_{\rho}^v(x_1)\phi_{\rho}^a(x_2)-\phi_{\rho}^a(x_1)\phi_{\rho}^v(x_2)\right]\right\}\cdot h(x_1,x_2,b_1,b_2),
\label{hlott}
\end{eqnarray}
where $C_F=4/3$, Sudakov factor $S_{\rho}(x_i;b_i;Q;\mu)$ and the hard function $h(x_1,x_2,b_1,b_2)$ can be found in Ref.~\cite{Zhang:2015mxa}.
The NLO space-like hard kernels can then be written in the form of
\begin{eqnarray}
H_{\rm LL22}^{(1)}(x_i,k_{iT},\mu, \mu_f, Q^2)=\mathcal{F}_{\rm LL22}^{(1)}(x_i,k_{iT},\mu, \mu_f, Q^2) \cdot H_{\rm LL22}^{(0)}(x_i,k_{iT},Q^2),
\label{eq:hll22}
\end{eqnarray}
\begin{eqnarray}
H_{\rm LT23}^{(1)}(x_i,k_{iT},\mu, \mu_f, Q^2)=\mathcal{F}_{\rm LT23}^{(1)}(x_i,k_{iT},\mu, \mu_f, Q^2) \cdot H_{\rm LT23}^{(0)}(x_i,k_{iT},Q^2),
\label{eq:hlt23}
\end{eqnarray}
\begin{eqnarray}
H_{\rm TL23}^{(1)}(x_i,k_{iT},\mu, \mu_f, Q^2)=\mathcal{F}_{\rm TL23}^{(1)}(x_i,k_{iT},\mu, \mu_f, Q^2) \cdot H_{\rm TL23}^{(0)}(x_i,k_{iT},Q^2),
\label{eq:htl23}
\end{eqnarray}
where $H^{(0)}$($H^{(1)}$) is the LO (NLO) hard kernel.
The explicit expressions of relevant correction functions $\mathcal{F}_{\rm LL22}^{(1)}$, $\mathcal{F}_{\rm LT23}^{(1)}$ and $\mathcal{F}_{\rm TL23}^{(1)}$ in Eqs.~(\ref{eq:hll22})-(\ref{eq:htl23}) are described as \cite{Zhang:2018bhj},
\begin{eqnarray}
 \mathcal{F}_{\rho, \rm LL22}^{(1)}(\mu, \mu_f, Q^2)
&=&\frac{\alpha_s(\mu_f) C_F}{4\pi} \left [ 
\frac{21}{4}\ln{\frac{\mu^2}{Q^2} -6\ln{\frac{\mu_f^2}{Q^2}}} + \frac{27}{8}\ln{x_1}\ln{x_2}\right. \non
&& - \frac{13}{8} \ln{x_1} + \frac{31}{16}\ln{x_2} -\ln^2\delta_{12}+\frac{17}{4}\ln x_1\ln\delta_{12}\non
&& \left. + \frac{23}{8}\ln\delta_{12}- \frac{17}{4}\ln^2{x_1} + \frac{1}{2} \ln2 + \frac{53}{4} + \frac{5}{48}\pi^2 \right ],
\label{eq:fll22}
\end{eqnarray}
\begin{eqnarray}
 \mathcal{F}_{\rho,\rm LT23}^{(1)}(\mu, \mu_f, Q^2)
&=& \frac{\alpha_s(\mu_f) C_F}{4\pi}\left  [\frac{21}{4}\ln{\frac{\mu^2}{Q^2} -6\ln{\frac{\mu_f^2}{Q^2}}} + \frac{9}{4}\ln{x_1}\ln{x_2} - \frac{21}{8}-2\ln{x_2}-\frac{1}{2}\ln^2{x_1} \right. \non
&&-\ln^2{x_2}\left. +\frac{17}{4} \ln{\delta_{12}} + \frac{31}{8} + \frac{17}{48}\pi^2 \right ],
\label{eq:flt23}
\end{eqnarray}
\begin{eqnarray}
\mathcal{F}_{\rho,\rm TL23}^{(1)}(\mu, \mu_f, Q^2)
&=& \frac{\alpha_s(\mu_f) C_F}{8\pi} \left [\frac{21}{2}\ln{\frac{\mu^2}{Q^2} -8\ln{\frac{\mu_f^2}{Q^2}}} + \frac{9}{4}\ln{x_1}\ln{x_2}
- \frac{3}{4} \ln^2{x_1} - \ln^2{x_2} \right. \non
&& - \frac{67}{8}\ln{x_1} - 2 \ln{x_2}  \left. + \frac{37}{8} \ln{\delta_{12}} + \frac{107}{8} - \frac{\pi^2}{3} \right ].
\label{eq:ftl23}
\end{eqnarray}
The subscripts $I, J = L, T$ $(k, l = 2, 3)$ of the term $\mathcal{F}_{\rho,IJkl}^{(1)}$ denote the different polarization components and twists from the initial and the final $\rho$ meson DAs.
For different polarizations of the initial and final $\rho$ meson, the contributions arise from different twist power of the initial and final DAs~\cite{Zhang:2015vor}.
The NLO correction to $F_{\rm LL}(Q^2)$ and $F_{\rm TT}(Q^2)$ with the initial and final states taking the twist-3 DAs shows similar power-law behavior with that combining twist-2 with twist-4 DAs~\cite{Cheng:2017vna}.
Thus we neglect the $F_{\rm LL33}(Q^2)$ and $F_{\rm TT33}(Q^2)$ terms in space-like form factor~\cite{Zhang:2018bhj}.

The LO analysis shows that the time-like hard kernel can be obtained from the space-like one by simple space transfer:$-Q^2\rightarrow Q^2$~\cite{Zhang:2022auc}.
The NLO correction to the time-like form factors are also derived from those to space-like one by suitable analytic continuation from $-Q^2\rightarrow Q^2$ under the follow relations,
\begin{eqnarray}
\ln(-Q^2-i\varepsilon)&=&\ln(Q^2)-i\pi,\non
\ln(-x_1Q^2+k_{1T}+i\varepsilon)&=&\ln(-x_1Q^2+k_{1T})+i\pi\Theta(-x_1Q^2+k_{1T}),\non
\ln(-x_1x_2Q^2+k_{1T}+i\varepsilon)&=&\ln(-x_1x_2Q^2+k_{1T})+i\pi\Theta(-x_1x_2Q^2+k_{1T}).
\label{eq:q2}
\end{eqnarray}
 The NLO time-like correction functions $\tilde{\mathcal{F}}_{\rm LL22}$, $\tilde{\mathcal{F}}_{\rm LT23}$ and $\tilde{\mathcal{F}}_{\rm TL23}$ in the parameter $b_i$ by the Fourier transformation from the transverse momentum space $k_{iT}$ to $b_i$ space take the form of:
\begin{eqnarray}
 \tilde{\mathcal{F}}_{\rho,\rm LL22}^{(1)}(\mu, \mu_f, Q^2)
&=& \frac{\alpha_s(\mu_f) C_F}{4\pi}\left [ \frac{21}{4}\ln{\frac{\mu^2}{Q^2}} 
+\left ( \frac{27}{16}\ln{x_2}+\frac{33}{8}\gamma_{E}-\frac{13}{16} +\frac{83}{16}i\pi \right )\ln{\frac{4x_1}{Q^2b_1^2}}   \right. \non
&&-6\ln{\frac{\mu_f^2}{Q^2}} -\frac{1}{4}\ln^2{\frac{4x_1x_2}{Q^2b_1^2}}+\frac{17}{16}\ln{\frac{4x_1}{Q^2b_1^2}}\ln{\frac{4x_1x_2}{Q^2b_1^2}}
-\frac{17}{16}\ln^2{\frac{4x_1}{Q^2b_1^2}} \non
&&+\left ( \frac{7}{16}i\pi-\frac{9}{8}\gamma_{E}-\frac{11}{16} \right ) \ln{\frac{4x_1x_2}{Q^2b_1^2}}
-\left (\frac{43}{16}i\pi+\frac{27}{8}\gamma_{E}-\frac{31}{16}\right )\ln{x_2}\non
&&+\left. \left (\frac{27}{16}+\frac{5}{4}\gamma_{E} \right )i\pi+\frac{17}{4}\gamma^2_{E}
+\frac{3}{8}\gamma_{E}-\frac{105}{24}\pi^2+\frac{1}{2}\ln2+\frac{53}{4} \right ],\label{eq:48a}
\end{eqnarray}
\begin{eqnarray}
 \tilde{\mathcal{F}}_{\rho,\rm LT23}^{(1)}(\mu, \mu_f, Q^2)
&=& \frac{\alpha_s(\mu_f) C_F}{4\pi}\left [\frac{21}{4}\ln{\frac{\mu^2}{Q^2}}
-6\ln{\frac{\mu_f^2}{Q^2}}-\frac{1}{8}\ln^2{\frac{4x_1}{Q^2b_1^2}}+\left (\frac{9}{8}\ln{x_2}+i\pi  \right. \right. \non
&&\left. -\frac{21}{6}-\gamma_{E}\right ) \ln{\frac{4x_1}{Q^2b_1^2}}-\left (\frac{27}{8}i\pi+\frac{9}{4}\gamma_{E}+2 \right )\ln{x_2}
+\frac{17}{8}\ln{\frac{4x_1x_2}{Q^2b_1^2}}\non
&&\left. -\ln^2{x_2}+\left (\frac{17}{16}-\frac{3}{2}\gamma_{E}\right )i\pi-\frac{1}{2}\gamma^2_{E}-
\frac{13}{8}\gamma_{E}+\frac{53}{48}\pi^2+\frac{31}{8} \right ],       \label{eq:48b}
\end{eqnarray}
\begin{eqnarray}
 \tilde{\mathcal{F}}_{\rho,\rm TL23}^{(1)}(\mu, \mu_f, Q^2)
&=& \frac{\alpha_s(\mu_f) C_F}{4\pi} \left [ \frac{21}{4}\ln{\frac{\mu^2}{Q^2}}
-8\ln{\frac{\mu_f^2}{Q^2}}-\left (\frac{9}{8}\ln{x_2}+\frac{3}{4}\gamma_{E}-\frac{67}{16} +\frac{3}{2}i\pi \right )\ln{\frac{4x_1}{Q^2b_1^2}} \right. \non
&&-\frac{3}{16}\ln^2{\frac{4x_1}{Q^2b_1^2}}+\frac{37}{16}\ln{\frac{4x_1x_2}{Q^2b_1^2}}
-\left (\frac{9}{8}i\pi+\frac{9}{4}\gamma_{E}-2 \right )\ln{x_2}\non
&&\left. -\ln^2{x_2}+\left (\frac{65}{8}-\frac{3}{4}\gamma_{E} \right )i\pi-\frac{3}{4}\gamma^2_{E}
+\frac{30}{8}\gamma_{E}+\frac{83}{48}\pi^2+\frac{107}{8} \right ],
\label{eq:48c}
\end{eqnarray}
where $\gamma_{E}$ is the Euler constant.

We can derive the double-b convolution of NLO time-like helicity amplitudes by Fourier transformation from transversal momentum space to the conjugate-parameter space,
\begin{eqnarray}
F_{\rm LL}(Q^2)&=&-\frac{32}{3}\pi C_F Q^2 \alpha_s(\mu)\int_0^1dx_1dx_2 \int_0^{\infty}b_1db_1b_2db_2 \cdot \exp[-S_{\rho}(x_i;b_i;Q;\mu)]\nonumber\\
      && \cdot \left \{  \left [-x_1+\frac{1}{2}\gamma_{\rho}^2(1-x_1)\right ]\phi_{\rho}(x_1)\phi_{\rho}(x_2) \left [1+\tilde{\mathcal{F}}_{LL22}(\mu,\mu_{f},Q^2)\right ] \right. \non
       &&\left. -\gamma_{\rho}^2\phi_{\rho}^t(x_1)\phi_{\rho}^s(x_2)-2\gamma_{\rho}^2(1+x_1)\phi_{\rho}^s(x_1)\phi_{\rho}^s(x_2) \right\}\cdot h^{\prime}(x_1,x_2,b_1,b_2).
\label{eq:8}
\end{eqnarray}
Similarly, the other two helicity amplitudes can be expressed as,
\begin{eqnarray}
F_{\rm LT}(Q^2)&=&-\frac{32}{3}\pi C_F Q m_{\rho} \alpha_s(\mu)\int_0^1dx_1dx_2 \int_0^{\infty}b_1db_1b_2db_2 \cdot \exp[-S_{\rho}(x_i;b_i;Q;\mu)]\nonumber\\
      &&\cdot \left \{ \frac{1}{2}\phi_{\rho(x_1)}\left [\phi_{\rho}^v(x_2)+\phi_{\rho}^a(x_2)\right]\left[1+\tilde{\mathcal{F}}_{LT23}(\mu,\mu_{f},Q^2)\right] \right. \nonumber\\
&& +\phi_{\rho}^T(x_1)\phi_{\rho}^s(x_2)\left[1+\tilde{\mathcal{F}}_{TL23}(\mu,\mu_{f},Q^2)\right]\nonumber\\
       &&\left.+\frac{1}{2}x_1\left[\phi_{\rho}^v(x_1)+\phi_{\rho}^a(x_1)\right]\phi_{\rho}(x_2)\left[1+\tilde{\mathcal{F}}_{LT23}(\mu,\mu_{f},Q^2)\right] \right\}\cdot h^{\prime}(x_1,x_2,b_1,b_2),\non
\label{eq:9}
\end{eqnarray}
\begin{eqnarray}
F_{\rm TT}(Q^2)&=&\frac{32}{3}\pi C_F  m_{\rho}^2 \alpha_s(\mu)\int_0^1dx_1dx_2 \int_0^{\infty}b_1db_1b_2db_2 \cdot \exp[-S_{\rho}(x_i;b_i;Q;\mu)]\nonumber\\
      &&\cdot\left\{(1+x_1)\left[\phi_{\rho}^a(x_1)\phi_{\rho}^a(x_2)-\phi_{\rho}^v(x_1)\phi_{\rho}^v(x_2)\right]\right.\nonumber\\
       &&\left.+(1-x_1)\left[\phi_{\rho}^v(x_1)\phi_{\rho}^a(x_2)-\phi_{\rho}^a(x_1)\phi_{\rho}^v(x_2)\right]\right\}\cdot h^{\prime}(x_1,x_2,b_1,b_2),
\label{eq:10}
\end{eqnarray}
The expressions of Sudakov factor $S_{\rho}(x_i;b_i;Q;\mu)$, which sums the large double logarithm on $k_T$ to all orders, has same modality with space-like form factor.
The hard function $h^{\prime}(x_1,x_2,b_1,b_2)$, which comes from the Fourier transfer of propagators on transversal components, can be obtained from space-like one $h(x_1,x_2,b_1,b_2)$ by substituting $x_1 \leftrightarrow -x_1$.
\begin{eqnarray}
h^{\prime}(x_1,x_2,b_1,b_2)= K_0(i\sqrt{x_1x_2}Q b_2)[\Theta(b_1-b_2) I_0(i\sqrt{x_1}Q b_2) K_0(i\sqrt{x_1}Q b_1)+ b_1 \leftrightarrow b_2],
\label{eq:11}
\end{eqnarray}
where the $K_0$ and $I_0$ are the modified Bessel function.

Here, the renormalization and factorization scales are set to the hard scale $t$ on the $k_{T}$ factorization theorem,
\begin{align}
\mu=\mu_{f}=t=\max(\sqrt{x_1}Q,\sqrt{x_2}Q,1/b_1,1/b_2).
\label{eq:12}
\end{align}

\section{Numerical analysis}
The numerical analysis is performed in this section, for which we adopt the standard two-loop QCD running coupling constant $\alpha_s(\mu)$ with the QCD scale $\Lambda_{\text{QCD}}=0.2857 \text{GeV}$.
We compute the LO and NLO contributions to the time-like $\rho$ meson EM form factors at leading and sub-leading twists via Eqs.~(\ref{eq:8})-(\ref{eq:10}) with the renormalization and factorization scale $\mu$ being set to the virtuality of the internal quark $\mu=max(\sqrt{x_i}Q,1/b_i)$.

\begin{figure}[htbp]
\begin{center}
\vspace{0.4cm}
\includegraphics[width=0.49\textwidth]{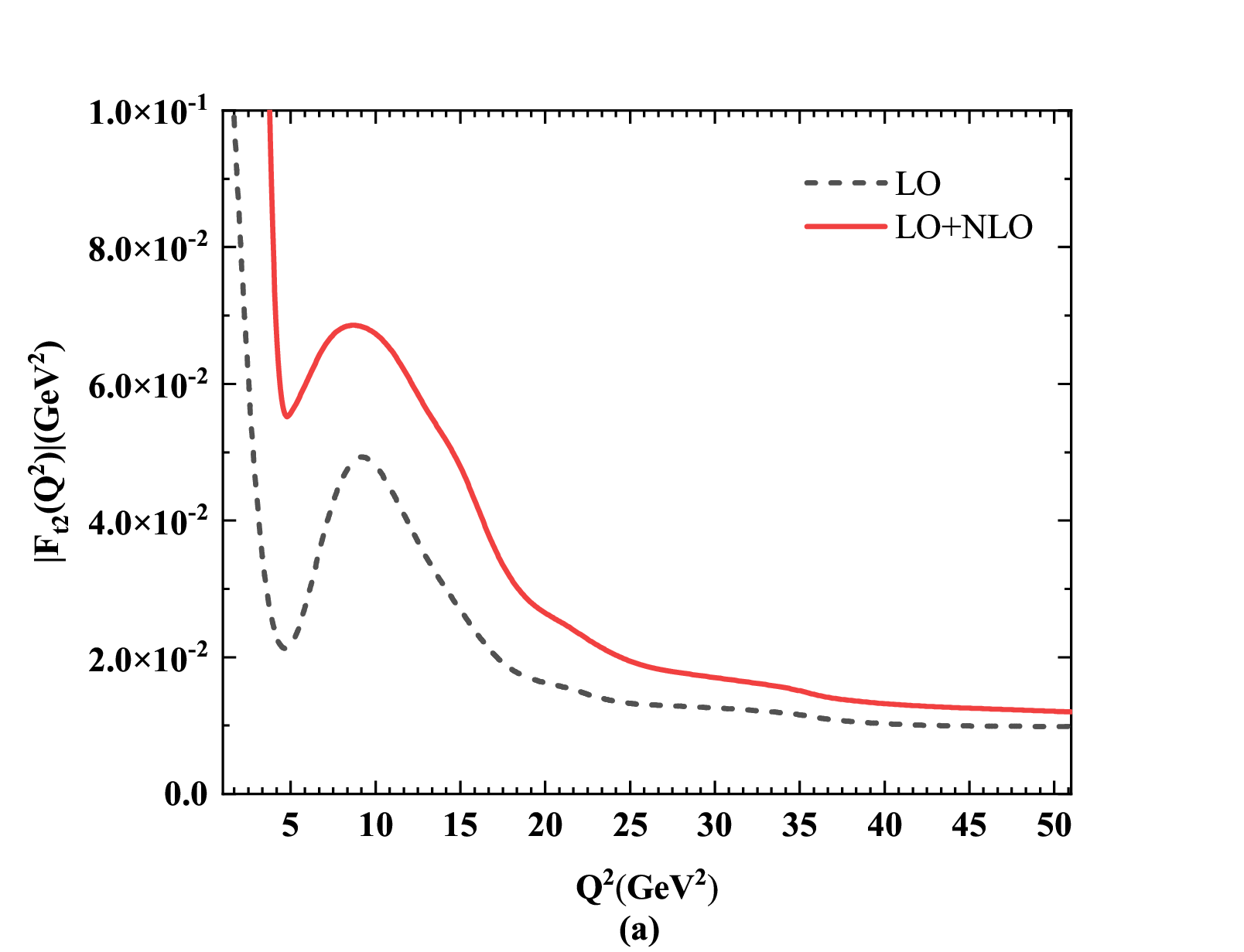}
\includegraphics[width=0.49\textwidth]{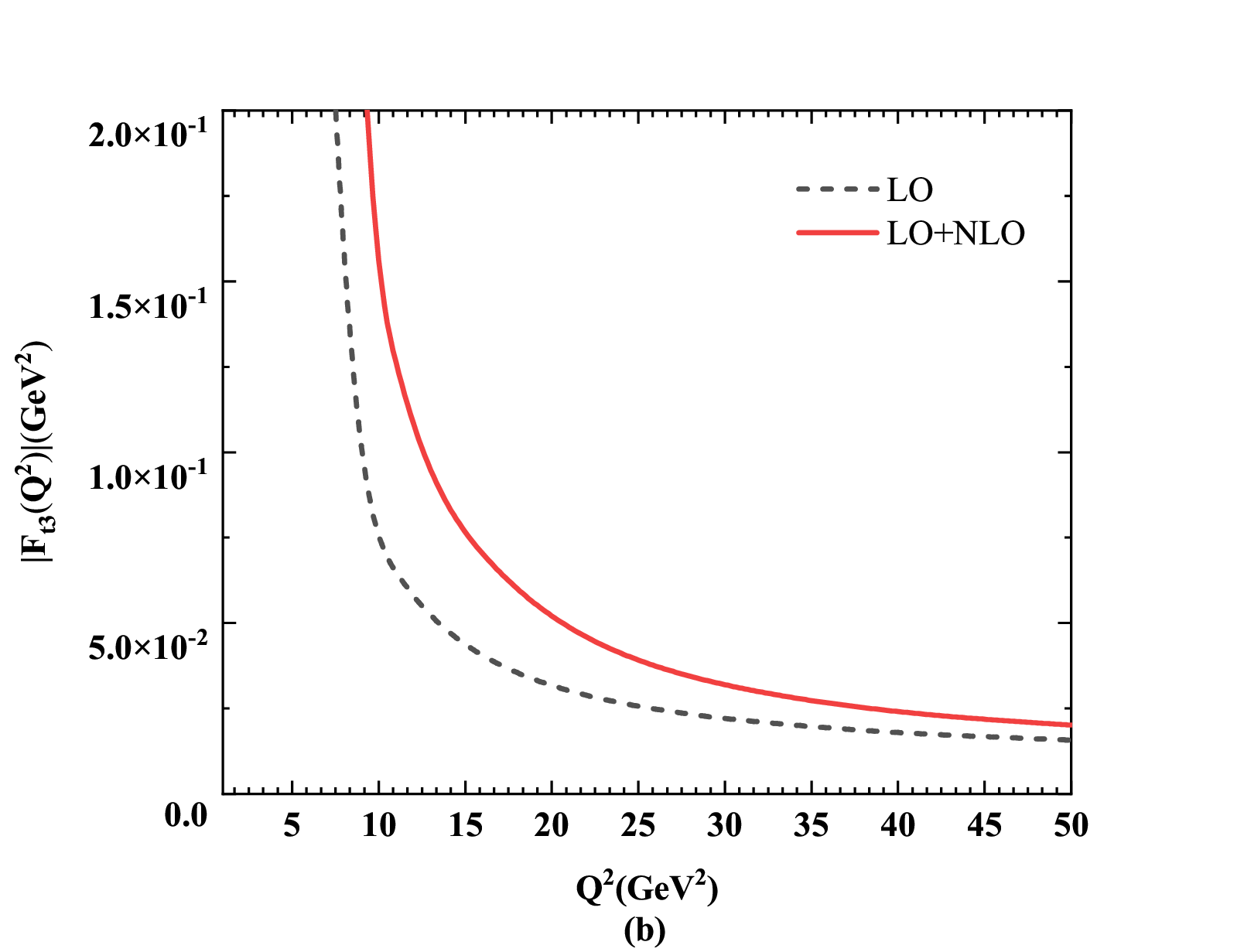}
\vspace{-0.4cm}
\caption{The PQCD predictions of the time-like EM form factor induced by the twist-2 (a) , and twist-3 (b) DAs at LO and NLO.}
\label{fig:ftform}
\end{center}
\end{figure}

The behaviors of the absolute value for $F_{t2}(Q^2)$ and $F_{t3}(Q^2)$ are shown in Fig.~\ref{fig:ftform} for $Q^2 < 50\,{\rm GeV}^2$, where $F_{t2(t3)}(Q^2)$ represents the sum of the EM form factors associated with twist 2 (twist 3).
In Fig.~\ref{fig:ftform}(a), the LO and NLO contributions at twist 2 reflect the oscillatory nature in the b space for $Q^2 \leq 20\,{\rm GeV}^2 $.
The LO time-like form factor exhibits an asymptotic magnitude, $|F_{t2}(Q^2)|\approx 0.011~\text{GeV}^2$ at large $Q^2$.
The NLO contribution to the time-like $\rho$ meson EM form factor decreases with $Q^2$ as expected in PQCD.
The NLO twist-2 correction to the magnitude is around $30\%$ at $Q^2=30 \text{GeV}^2$, and less than $20 \%$ for $Q^2 >50 \text{GeV}^2$ in comparison with LO result.
The NLO twist-3 correction to the magnitude of the LO twist-3 contribution is smaller than $30\%$ in the region $Q^2 > 30 \text{GeV}^2$, shown in Fig.~\ref{fig:ftform}(b).
It is worth mentioning that the $\rho$ EM form factor is dominated by twist-3 contribution instead of by twist-2 one because of the end-point enhancement.
This enhancement is understood easily as follows: the virtual quark and gluon propagators behave like $1/x_1$ and $1/(x_1x_2)$, respectively.
The twist-2 $\rho$ DAs are proportional to $O(x)$, but the twist-3 DAs remain constant $O(1)$ at small $x$, which then enhance the end-point contribution dramatically.
This enhancement was also observed in perturbative evaluation of the $B \to \pi$ transition form factor \cite{Kurimoto:2001zj} and $\pi$ meson EM form factor \cite{Hu:2012cp,Cheng:2015qra} and the recent $\Lambda_b \to p$ transition form factor \cite{Han:2022srw,Yu:2024cjd}.

\begin{figure}[htbp]
\begin{center}
\vspace{0.4cm}
\includegraphics[width=0.8\textwidth]{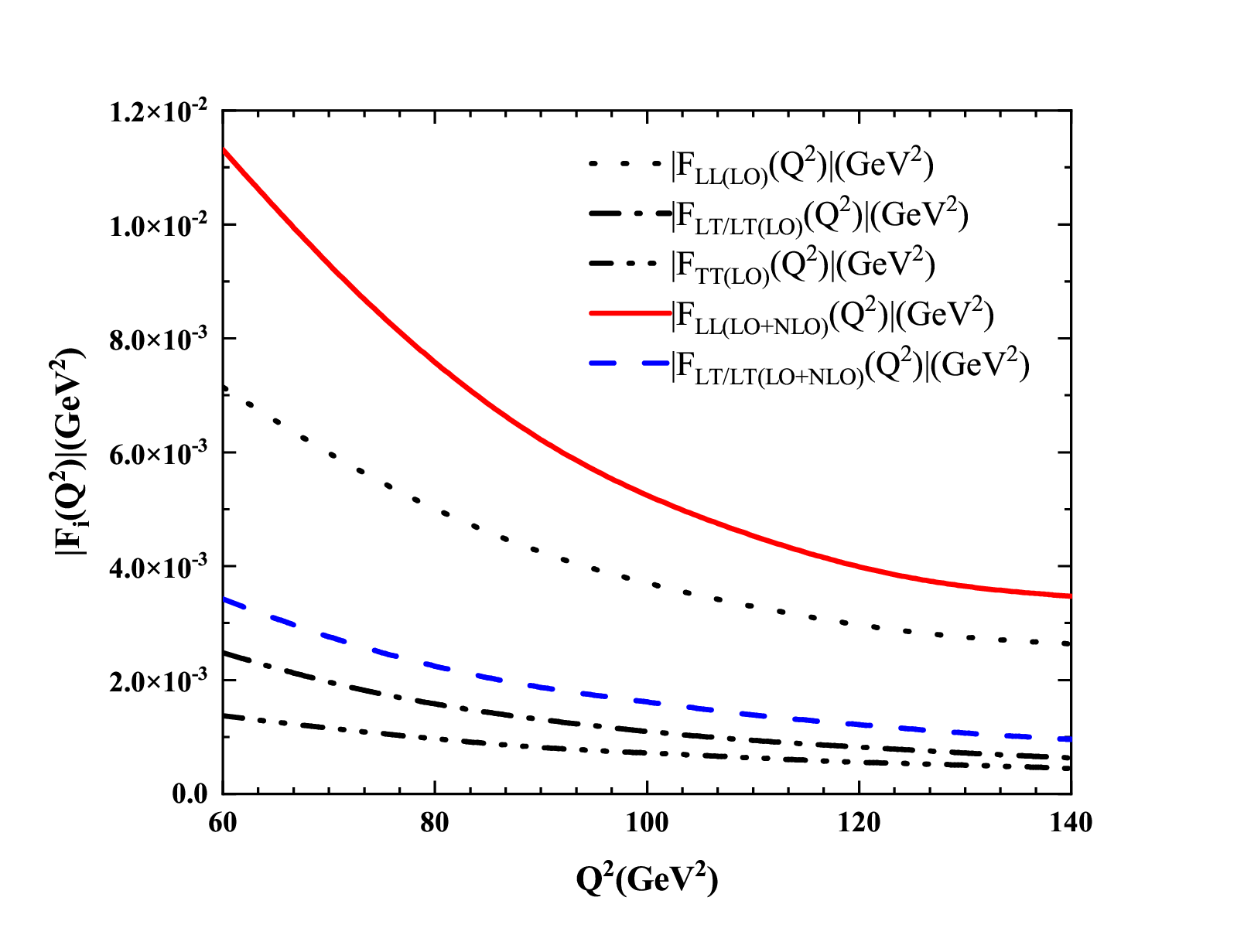}
\vspace{-0.4cm}
\caption{The PQCD predictions of $\rho$-meson helicity amplitudes $\lvert F_i(Q^2)\rvert$  with $i=\rm LL,LT/TL,TT$ for $Q^2$ dependence up to NLO.}
\label{fig:f3form}
\end{center}
\end{figure}

In Fig.~\ref{fig:f3form}, we plot the dependence of the absolute value for three helicity amplitudes on the considered invariant mass squared $Q^2$ in the region of $60~\text{GeV}^2<Q^2<140~\text{GeV}^2$.
Each helicity amplitude approaches an asymptotic value at large $Q^2$.
The NLO contribution to the helicity amplitude is also displayed in Fig.~\ref{fig:f3form}, which decreases with $Q^2$ as expected in
PQCD.
Compared to the LO result, the NLO correction to the $|F_{\text{LL}}(Q^2)|$ associated with longitudinal polarization is about $40\%$ at $Q^2= 100\text{GeV}^2$, and less than $30\%$ for $Q^2>140 \text{GeV}^2$.
For $|F_{\text{LT/TL}}(Q^2)|$, the NLO contribution is about $30\%$ in the whole region of $Q^2$.
While for the $|F_{\text{TT}}(Q^2)|$, we can only calculate the LO contribution at present and find the LO value is about five times smaller than that of $|F_{\text{LL}}(Q^2)|$ at large $Q^2$.

Recently, the ratio of the moduli squared of three independent helicity amplitudes at $\sqrt{s}=10.58 \text{GeV}$ has been reported by BABAR Collaboration \cite{BaBar:2008fsh}
{\small
\beq
|F_{\rm LL}|^2:|F_{\rm LT/TL}|^2:|F_{\rm TT}|^2=(0.51\pm0.14\pm0.01):(0.10\pm0.04\pm0.01):(0.04\pm0.03\pm0.01),
\label{eq:13}
\eeq}
with the following normalization $|F_{LL}|^2+4|F_{LT/TL}|^2+2|F_{TT}|^2=1$.
%Taking the central value $0.51$ for $|F_{LL}|^2$ with this normalization, the %theoretical ratio yields $|F_{LL}|^2:|F_{LT/TL}|^2:|F_{TT}|^2=0.51:(1.1\times10^{-2}):(1.4\times10^{-5})$ in \cite{deMelo:2016lwr}.
We calculate the moduli of three helicity amplitudes in Eqs.~(\ref{eq:8})-(\ref{eq:10}) using the same energy as that of BABAR experiment, then the theoretical ratio:
\begin{eqnarray}
|F_{\text{LL}}|^2:|F_{\text{LT/TL}}|^2:|F_{\text{TT}}|^2&=& \left\{\begin{array}{ll}
0.436:0.117:0.082,
& (\text{LO}),\\
0.522:0.107:0.025,
& (\text{NLO}).\\
\end{array} \right.
 \label{eq:hamiltonian01}
\end{eqnarray}
As mentioned above, the NLO correction enhanced the longitudinal helicity amplitude largely and makes slightly influence on the transverse helicity amplitude.
Our predictions are consistent with the measurement within errors at NLO level, indicating that the asymptotic region has been reached at $\sqrt{s}=10.58 \text{GeV}$ in the BABAR experiment \cite{BaBar:2008fsh}.

\begin{figure}[htbp]
\begin{center}
\vspace{0.4cm}
\includegraphics[width=0.8\textwidth]{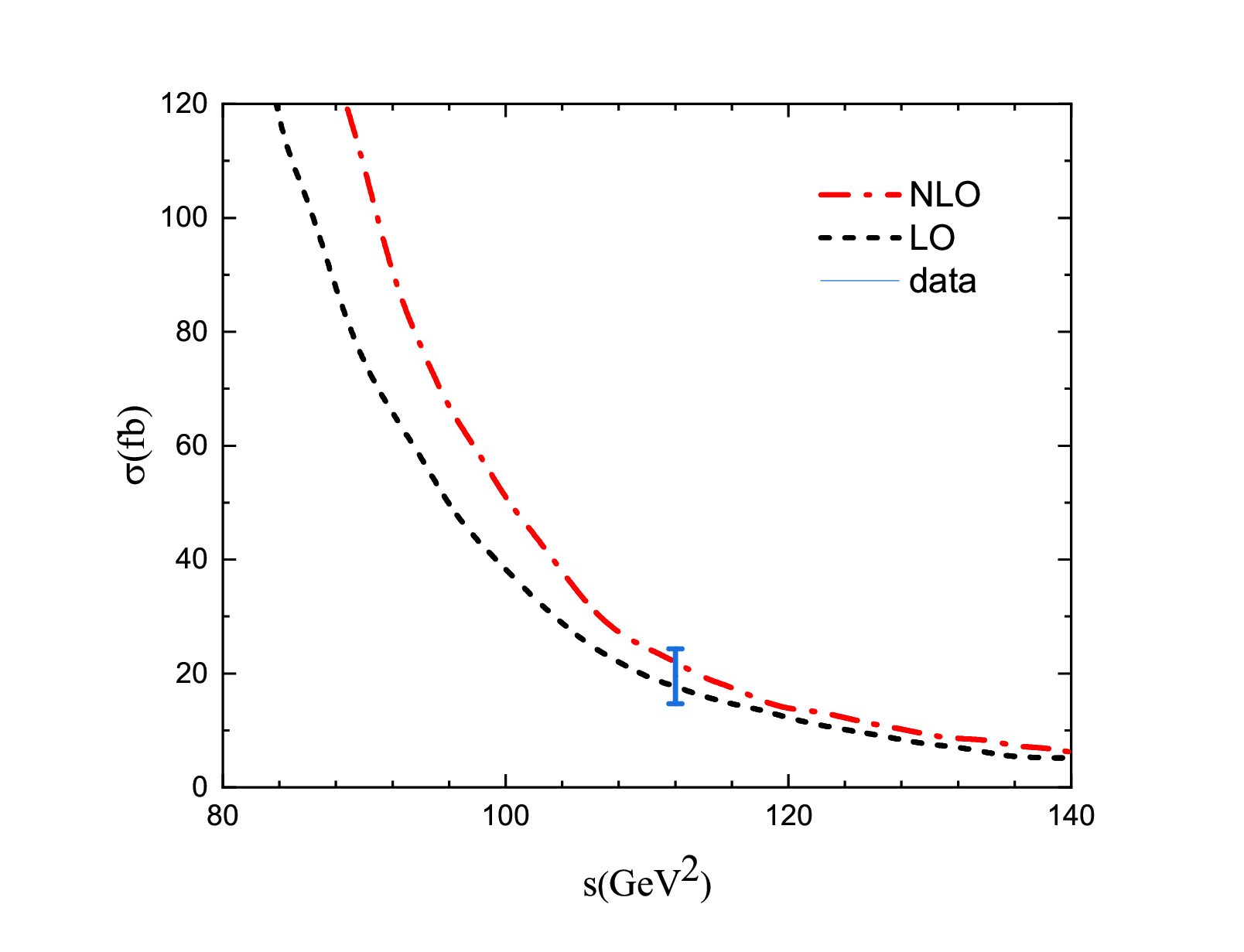}
\vspace{-0.5cm}
\caption{The cross-section of $e^+e^-\rightarrow \rho^+\rho^-$ as a function of s and the BABAR data at $s=112~\text{GeV}^2$ \cite{BaBar:2008fsh}.}
\label{fig:fig4}
\end{center}
\end{figure}

The cross-section can be written in terms of the time-like form factors:
\begin{align}
\sigma=\frac{\pi\alpha^2\beta^3}{3q^2}\frac{1}{4m_{\rho}^2(\eta-1)}(|{F}_{LL}|^2+4|{F}_{LT/TL}|^2+2|{F}_{TT}|^2),
\label{eq:15}
\end{align}
where $\alpha = 1/137$ is the electromagnetic constant, $\beta= \sqrt{1-4m_{\rho}^2/q^2}$ is the $\rho$-meson velocity, $\eta=Q^2/4m_{\rho}^2$.
The cross-section of $e^+e^-\rightarrow \rho^+\rho^-$ up to NLO as a function of $s=Q^2$ is shown in Fig.~\ref{fig:fig4} in the range of $s=[80~{\rm GeV}^2,  140~{\rm GeV}^2]$.
The total cross-section for $e^+e^-\rightarrow \rho^+\rho^-$ was measured to be $\sigma=19.5\pm1.6({\rm stat})\pm3.2{\rm (syst)\,}$ fb by BABAR collaboration near $\sqrt{s}=10.58$ GeV \cite{BaBar:2008fsh}.
For comparison with experimental data, the value of total cross-section near $s=112\,{\rm GeV}^2$ is proposed.
Compare to $\sigma=17.4$ fb from the LO contribution in \cite{Zhang:2022auc}, the value is roughly $21.2$ fb up to the NLO.
Our prediction are in agreement with the experimental data within uncertainties.

\begin{figure}[htbp]
\begin{center}
\vspace{0.4cm}
\includegraphics[width=0.8\textwidth]{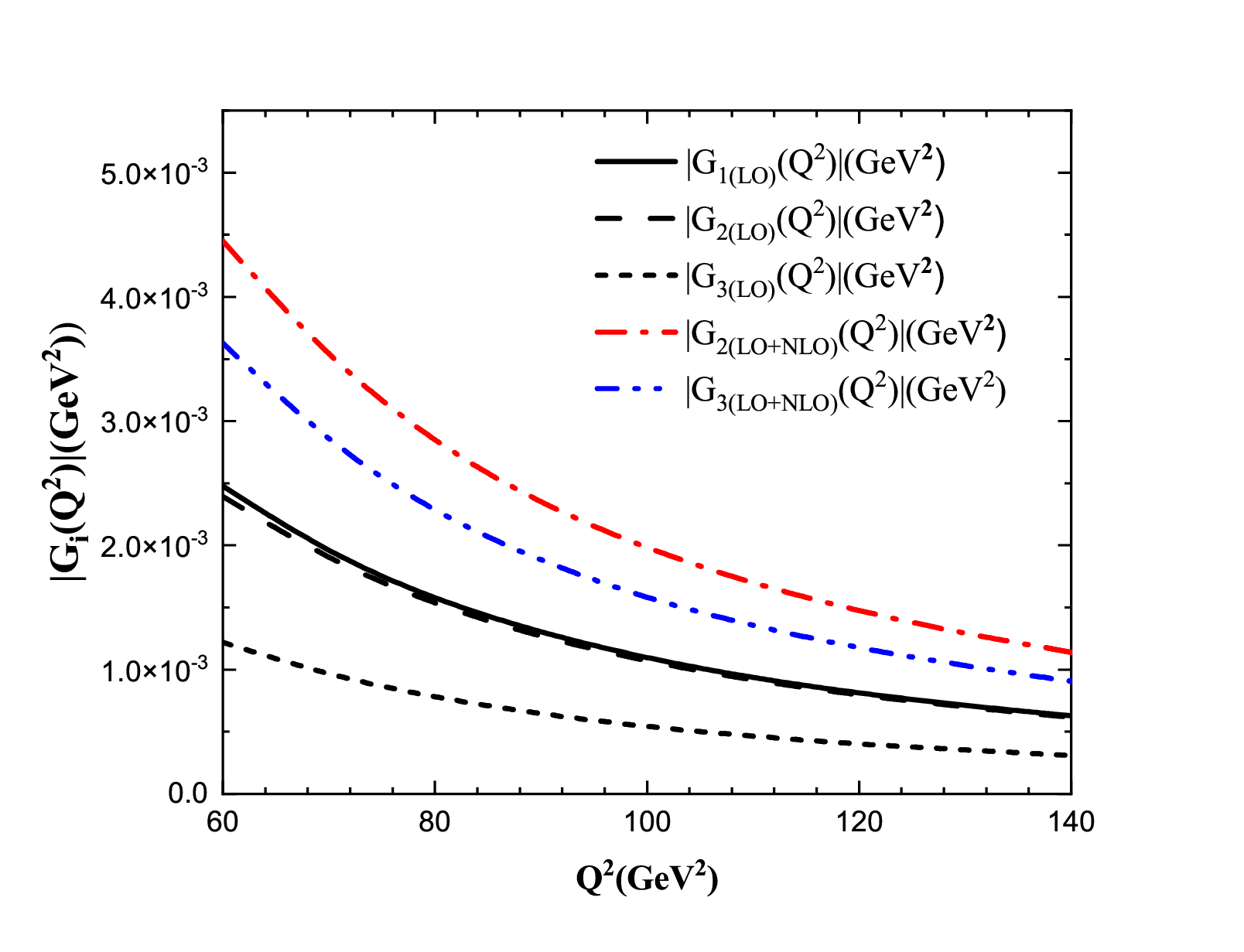}
\vspace{-0.4cm}
\caption{The PQCD predictions for the dependence of time-like form factor $\lvert G_i(Q^2)\rvert$ on $Q^2$, with $i=1,2,3$.}
\label{fig:gform}
\end{center}
\end{figure}

The results for $\rho$ meson charge $G_1(Q^2)$, magnetic $G_2(Q^2)$, and quadrupole $G_3(Q^2)$  form factors can be evaluated easily based on Eq.~(\ref{fform}).
We show the behavior for the $Q^2$ dependence of the time-like form factor $G_i(Q^2)$ in the region of $60~\text{GeV}^2< Q^2 <140~\text{GeV}^2$ in Fig.~\ref{fig:gform}.
The NLO correction can enhance the magnitude of LO form factor $|G_3(Q^2)|$ by 40$\%$ in the considered invariant mass squared $Q^2> 100\text{GeV}^2$. 
It is shown that the magnitude of LO form factors $|G_1(Q^2)|$ is the same as that of $|G_2(Q^2)|$.
With the inclusion of NLO correction, the value of form factors $|G_2(Q^2)|$ becomes sizable.
We can estimate the ratio of three form factors at large $Q^2$ with the NLO correction, $G_1:G_2:G_3=0.5:2:1$.
In Refs.~\cite{deMelo:1997hh,Hawes:1998bz,Roberts:2011wy,Bakker:2002mt}, there have been a few theoretical computations that reported a zero in the space-like region of $G_1$.
Space and time-like form factors in the asymptotic momentum transfer region should be identical to the Phragm\`{e}n-Lindel\"off theorem.
As the form factors are analytic functions of the momentum transfer $Q^2$, they should satisfy this theorem.
Thus, the value of time-like region for $G_1$ close to a zero is expected, which supports the tiny value of the time-like $G_1$ in this work.
Our prediction of ratio is satisfied with universal ratios in Eq.~(\ref{3ffratio}).

\section{Conclusions}
In this work, we have calculated the time-like $\rho$ meson EM form factors up to NLO in the $k_T$ factorization formalism.
Leading-twist together with sub-leading twist contributions are included.
It was observed that the NLO corrections in magnitude change the LO leading-twist results by roughly $30\%$ for the EM form factor.
While for the LO sub-leading twist, the NLO contribution is less than $30\%$ at $Q^2>30~\text{GeV}^2$.
It has been realized that the $\rho$ meson EM form factor is dominated by sub-leading contribution instead of by leading one because of the end-point enhancement.
We predicted the ratio of the moduli of three helicity amplitudes and total cross section at $\sqrt{s}=10.58$ GeV by including sub-leading twists at the NLO level, which are consistent with measurements from BABAR Collaboration.
Stimulated by our work, we have the confidence on computing these complex time-like form factors directly in the PQCD approach, which can be adopted in multi-body hadronic $B$ meson decays.

\section{ACKNOWLEDGEMENTS}
This work is supported by the National Natural Science
Foundation of China under Grants No.12005103, 11775117 and 11847141, and also by the Practice Innovation Program of Jiangsu Province under Grant No. KYCX18-1184 and Natural Science Research program of Huaian  under Grant No. HABZ202314.

%%---------------------------------------------------------------------------------------

\end{document}